\documentclass[12pt]{spieman} 
\usepackage[]{graphicx}
\usepackage{setspace}
\usepackage{tocloft}

\title{Modeling of Hybridized IR Arrays for Characterization of Interpixel Capacitive Coupling} 

\author{Kevan Donlon\supscr{a}, Zoran Ninkov\supscr{a}, Stefi Baum\supscr{a,b}, Linpeng Cheng\supscr{a}}

\affiliation{\supscrsm{a}Rochester Institute of Technology, Chester F. Carlson Center for Imaging Science, 54 Lomb Memorial Drive, Rochester, New York, 14623\\
\supscrsm{b}University of Manitoba, Department of Physics and Astronomy, 66 Chancellors Cir, Winnipeg, MB, Canada}

\cftpagenumbersoff{figure}
\cftpagenumbersoff{table} 
\begin{document} 
\maketitle 

\begin{abstract}
Inter pixel capacitance (IPC) is a deterministic electronic coupling resulting in a portion of signal incident on one pixel of a hybridized detector array being measured in adjacent pixels. Data collected by light sensitive HgCdTe arrays which exhibit this coupling typically goes uncorrected or is corrected by treating the coupling as a fixed point spread function. Evidence suggests that this coupling is not uniform across signal and background levels. Sub-arrays of pixels using design parameters based upon HgCdTe indium hybridized arrays akin to those contained in the James Webb Space Telescope's NIRcam have been modeled from first principles using Lumerical DEVICE software. This software simultaneously solves Poisson's Equation and the Drift Diffusion Equations yielding charge distributions and electric fields. Modeling of this sort generates the local point spread function across a range of detector parameters. This results in predictive characterization of IPC across scene and device parameters that would permit proper photometric correction and signal restoration to the data. Additionally, the ability to visualize potential distributions and couplings as generated by the models yields insight that can be used to minimize IPC coupling in the design of future detectors.
\end{abstract}


{\noindent \footnotesize{\bf Address all correspondence to}: Kevan Donlon, Rochester Institute of Technology, Chester F. Carlson Center for Imaging Science, 54 Lomb Memorial Drive, Rochester, New York, 14623; E-mail: \linkable{KevanADonlon@gmail.com} }

\begin{spacing}{2} 

\section{Introduction}
\label{sect:intro} 
The goal of this work was to model and characterize, from first principles, the electronic interpixel capacitive (IPC) cross talk between adjacent pixels in a hybridized imaging array. This characterization was then compared to measurements obtained from analysis of single pixel cosmic ray event (CRE) data. The presence of this cross talk causes a spread in images of point sources.  Mathematically, this is characterized through convolution by a blur kernel.  This blurring also results in a reduction of apparent Poissonian noise\cite{Moore03} as well as reduction in accuracy of per pixel flux measurements.

IPC coupling was first observed in test structures fabricated by the Lawrence Berkeley National Laboratory (LBNL). These structures were designed to establish a set of measurements and methodology which could later be used to test the ATLAS or other silicon particle detectors for the Large Hadron Collider (LHC) or Tevatron colliders. These test structures were manufactured with 50 ${\mu}$m by 536 ${\mu}$m pixels\cite{Gorfine01}. The IPC of these LBNL test sensors was large due to the substantial area of overlap between such large pixels. For these arrays, electrostatic simulations were performed exclusively between adjacent metal contacts to determine IPC. This assumption resulted in an underestimate of the overall coupling\cite{Gorfine01}. This result indicates that there is a coupling in addition to the coupling between the metal contacts.

\begin{figure}
\begin{center}
\begin{tabular}{c}
\includegraphics[height=6.9cm]{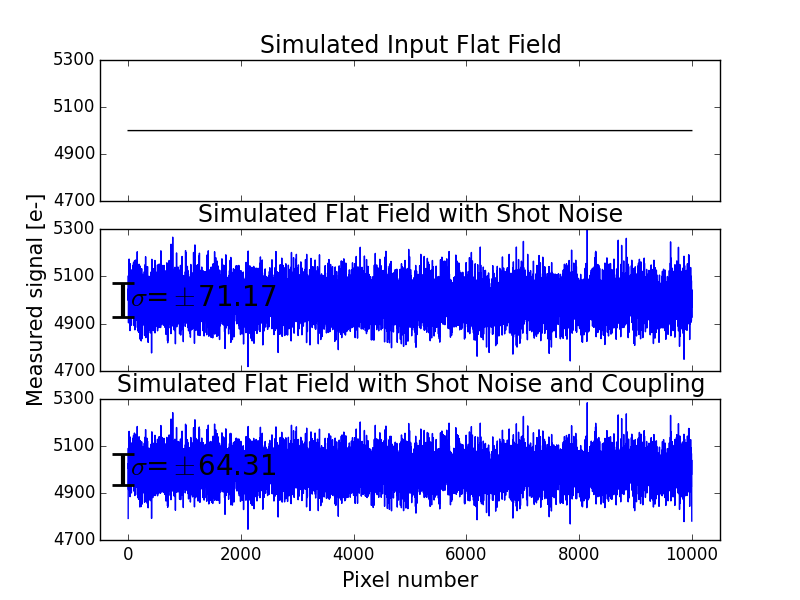}
\end{tabular}
\end{center}
\caption {(top) Flat field with a level of 5000 $e^-$ on a line array of 10000 pixels with no noise. (mid) Shot noise applied to the preceding flat field. (bot) Pixel value after IPC coupling of 0.05 to nearest neighbors. Measured variance has been reduced by 18\% from 5053 to 4135. } 
\label{fig:1}	
\end{figure} 
IPC was subsequently observed in hybrid Silicon PIN and InSb arrays designed for use in astronomy\cite{Moore03}. When examining the noise properties of these devices, IPC reduces the apparent Poissonian noise by acting as a smoothing filter, as illustrated in figure~\ref{fig:1}. This noise reduction invalidates key assumptions for gain and noise estimations made using the photon-transfer curve method. Assuming that the noise in a signal is exclusively the result of shot noise, a flat field will have variance equal to its signal strength in number of events\cite{Janesick01}. As a result, a plot of variance as a function of signal strength in ADU will have a slope equal to the inverse of the conversion gain [$e^-$/ADU]. However, when the shot noise is diminished by IPC the variance measurements are lower than expected resulting in an overestimation of conversion gain. This erroneous gain propagates into calculations of quantum efficiency resulting in an overestimation of relative quantum efficiency and detective quantum efficiency when coupling is as small as on the order of 1\% .\cite{Moore03}$^,$\cite{Moore06} IPC went unnoticed in the detector community for an extended period of time resulting in many sensors overestimating their quantum efficiencies.  Measurements of IPC were made in part due to this quantum efficiency overestimation and in part due to changes to the edge spread function of these arrays that could not be explained by diffusion\cite{Moore06}. IPC also serves to reduce the measured gain when using a fixed source (e.g. Fe-55) method of gain calibration by attenuating the expected signal\cite{Moore06}. This necessitates adoption of a more difficult direct capacitive comparison method\cite{Finger06} to accurately measure conversion gain.

Even on an array which used a direct capacitive comparison method to characterize conversion gain: the peak magnitude of a point source would be lower than expected, the spatial extent would be larger than expected, and any high spatial frequency fluctuations would be of lower magnitude. This results in lower resolving power and a smoothing of any edges and fine features.

After observation of this IPC effect was made, an explanation as to the mechanism of coupling was posited. In a detector with an unbiased bulk, coupling within the bulk region is approximately zero as the bulk material cannot support fields; this requires that the IPC coupling occur in the insulating layer\cite{Moore03}. In a detector operating with a strongly biased bulk IPC coupling can occur within the depletion region as well as in the insulating layer unless a field control grid is present which serves to minimize coupling within the insulating layer\cite{Moore03}.

\begin{figure}
\begin{center}
\begin{tabular}{c}
\includegraphics[height=6.3cm]{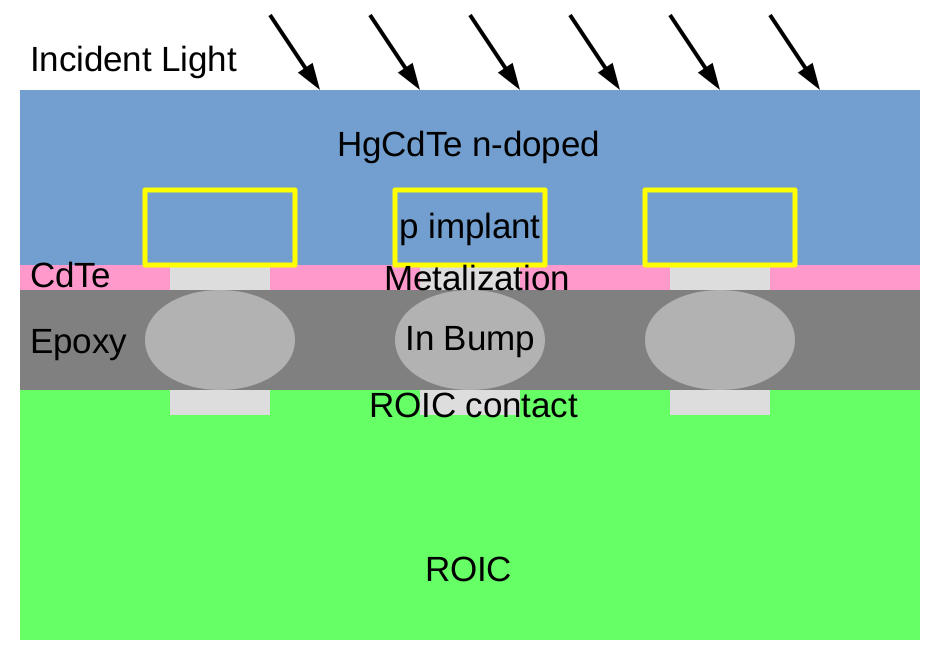}
\end{tabular}
\end{center}
\caption {Cross section of a HgCdTe Hybridized array showing indium bump bonding between a p-on-n structure photodiode array and ROIC (not to scale). } 
\label{fig:2}
\end{figure} 

The detector array on which the models for IPC presented in this paper are based is a photo-active HgCdTe photo-diode array with a 2.7${\mu}$m cutoff hybridized to an H2RG read-out integrated circuit (ROIC) with pixel pitch of 18${\mu}$m. Hybridization to connect the ROIC and detector layers is done through indium bump bonds. The detectors are assumed thinned for operation and were of n-type bulk. A sample cross-section of the detector array is presented in figure~\ref{fig:2}. The coupling was emulated by simulating the electrostatics and semi-conductor physics through iteratively solving for mutually consistent steady state solutions. A commercial-grade device simulator, Lumerical DEVICE, that self-consistently solves the Poisson and drift-diffusion equations was used to perform the calculations\cite{Lumeric00}. Simulations were made over an area of nine pixels arranged in a three by three grid. The ROIC itself was not simulated, resulting in the reported couplings occurring only in the sensor and hybridization layers.  Electrostatic potentials on the bulk and indium bumps were initialized which corresponded to a state of collected charge against a fixed reverse bias. These potentials and their corresponding fields mobilize charge carriers within the bulk. Interpixel capacitance can be calculated from this charge-voltage relationship.

This characterization has a two-fold intention. Primarily, it establishes a methodology by which the IPC coupling strength can be deduced. This includes characterization as a function of both physical properties of the sensor (dielectric strength of epoxy, implant size, temperature, doping profile, indium size, etc.) as well as properties of the scene (signal intensity, background intensity, etc. ). These variations are then mathematically formalized to generate a local coupling kernel. Additionally, these models have predictive power. Observation of the fields and charge distributions can give insight into the mechanisms of coupling. Optimization in the models to minimize IPC can influence sensor design for future detector array manufacturing.

An agreement is observed between collected data using CREs and the simulated data generated with the model described here. This validates the model as capable of generating realistic couplings. 

\section{Fundamental Theory}
\label{sect:Fund_Theo}
\subsection{Circuit mathematics}
From a theoretical perspective, the goal is to establish the magnitude of the capacitance between one pixel and its neighbors. Capacitance between two objects is defined by:
\begin{equation}
\label{eq:cap}
C = \frac{Q}{\Phi} \ ,
\end{equation}
Where a local charge distribution (Q) induces an electrostatic potential (${\Phi}$) or due to reciprocity a local potential induces a charge distribution.

\begin{figure}
\begin{center}
\begin{tabular}{c}
\includegraphics[height=4.1cm]{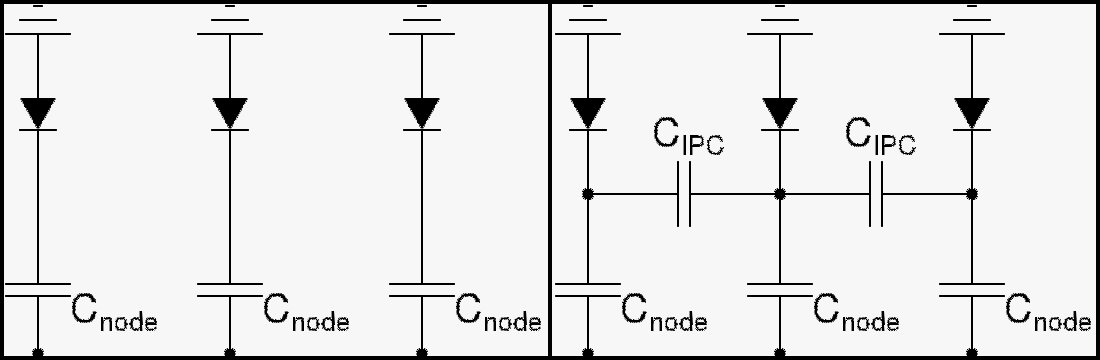}
\end{tabular}
\end{center}
\caption {From side circuit diagram of three pixels in a line with (left) no coupling and (right) nearest neighbor coupling. } 
\label{fig:3}
\end{figure}

\begin{figure}
\begin{center}
\begin{tabular}{c}
\includegraphics[height=6.35cm]{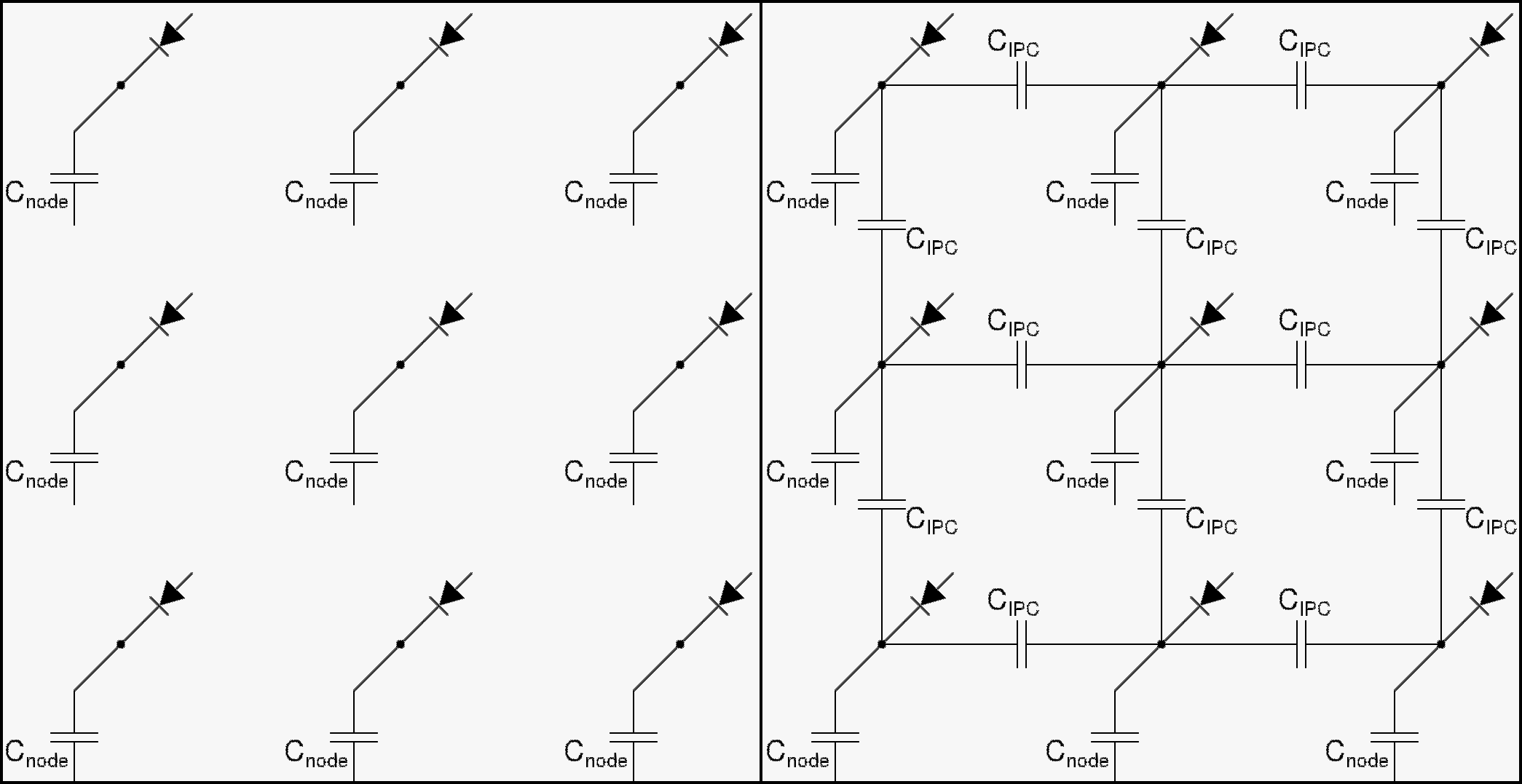}
\end{tabular}
\end{center}
\caption {From above circuit diagram of a three by three array with (left) no coupling and (right) nearest neighbor coupling. }
\label{fig:4} 
\end{figure}

Photodiodes convert incident photons to charge carriers. These charge carriers propagate through the detector producing a voltage at an output node which is available for digitization. The particular relationship between a charge collected in a pixel and the voltage read-out is referred to as the nodal-capacitance ($C_{node}$). In an ideal sensor this is the only capacitance to which the collected charge would be subject (figure~\ref{fig:3} and~\ref{fig:4} left). However, due to the geometric arrangement of sensor elements, an additional capacitance ($C_{IPC}$) is present which deterministically couples the signal from one pixel to its nearest neighbors\cite{Moore03}(figure~\ref{fig:3} and~\ref{fig:4} right).

Letting $Q_n$ and ${\Phi}_n$ be the charge and potential measured 'n' nodes from the pixel where generation occurs, yields a coupling coefficient ${\alpha}$ defined by:
\begin{equation}
\label{eq:coup}
{\alpha} \equiv \frac{{\Phi}_1}{{\Phi}_0} = \frac{Q_1}{Q_0}=\frac{C_{IPC}}{4C_{IPC}+C_{node}} \ .
\end{equation}
This result allows the coupling strength to be modeled on the output in analog digital units (ADU) as a convolution with the kernel\cite{Moore04}:
\begin{equation}
\label{eq:kernel}
K(m,n)= \left[ \begin{array}{ccc} 0 & {\alpha} & 0 \\ {\alpha} & 1-4{\alpha} & {\alpha} \\ 0 & {\alpha} & 0 \end{array} \right].\
\end{equation}
That is to say, when a signal generates a spatial charge distribution on pixel (i,j) of Q(i,j) [$e^-$] with a gain of g [ADU/$e^-$] the output pattern, in ADU, would be given by:
\begin{equation}
\label{eq:convolution}
S(i,j)=g Q(i,j) \ast K(m,n) \ ,
\end{equation}
where $\ast$ is the convolution operator and (m,n) are pixel position offsets from -1 to 1 to be integrated over by the convolution. The nature of this coupling parameter ($\alpha$) is not established to be of a particular formulaic type but evidence indicates that it is not exclusively the parallel cylinder capacitance between adjacent indium bumps as has been earlier posited \cite{Brown06}. $\alpha$ has instead been observed to also depend on scene parameters such as signal intensity, background intensity, temperature, etc. as well as intrinsic sensor parameters such as indium bump size, pixel pitch, etc.
\subsection{Modeling}
The Lumerical DEVICE software\cite{Lumeric00} used simulates the generation, movement, and interaction of charge carriers in the detector as governed by Poisson's equation:\cite{Ohanian07}
\begin{equation}
\label{eq:poisson}
\mathbf{\nabla}^2 {\Phi}=-4{\pi}{\rho}\ ,
\end{equation}
under the constraints of the drift diffusion or continuity equations:\cite{Sze07}
\begin{equation}
\label{eq:dde_1}
\frac{\mathbf{J_n}}{-q}=-D_n\mathbf{\nabla}n-n{{\mu}_n}\mathbf{E}\ ,
\end{equation}
\begin{equation}
\label{eq:dde_2}
\frac{\mathbf{J_p}}{q}=-D_p\mathbf{\nabla}p-p{{\mu}_p}\mathbf{E}\ ,
\end{equation}
\begin{equation}
\label{eq:dde_3}
\frac{{\partial}n}{{\partial}t}=\frac{-\mathbf{{\nabla}}\mathbf{J_n}}{-q}+R\ ,
\end{equation}
\begin{equation}
\label{eq:dde_4}
\frac{{\partial}p}{{\partial}t}=\frac{-\mathbf{{\nabla}}\mathbf{J_p}}{q}+R\ .
\end{equation}
These equations are solved iteratively utilizing Newton's method until a simultaneous consistent solution is found.

The end point of calculation represents the final steady state solution established after charges reach equilibrium. The initial configuration is defined by the spatial distribution of n and p carriers determined by doping profiles and concentrations; introduction of conductors with their associated potentials, work functions, and circuit connections; and any insulators/dielectrics in the intermediate volumes.

Taken in tandem equations ~\ref{eq:poisson} - ~\ref{eq:dde_4} resolve to a pair of coupled, second order in space, first order in time, differential equations with charge sources and sinks established by metal contacts through:
\begin{equation}
\label{eq:boundary_1}
\mathbf{\nabla}^2 {\Phi} = -4{\pi}(p-n)q \,
\end{equation}
\begin{equation}
\label{eq:boundary_2}
{\Phi}=V-\frac{W}{q} \,
\end{equation}
and recombination rates based on temperature and mobility related parameters.  In these equations p and n are carrier concentrations, V is the voltage applied, W is the work function of the conductor, and q is the fundamental charge unit. Taken as a set this formalism is known as the Van Roosbroeck model\cite{Antoni00}.

In our simulations the ROIC was not included. Due to the distance of the ROIC from the highest magnitude electric fields, their impact on the results presented here is expected to be minimal. The design of ROICs vary to a considerable degree, lacking any particular general design that could be incorporated into the model. Without first principle modeling of the ROIC, a nodal capacitance is not modeled. This nodal capacitance is required to determine the coupling strength as shown in equation \ref{eq:coup}. In order to generate coupling strengths ($\alpha$), the value of $C_{node}$ was taken to be 33.5 fF as this was measured by direct capacitive comparison in similar arrays\cite{Finger06}. This allows approximate coupling strengths, in percent of signal, to be calculated.

In our model a potential difference is initially set across the diode. This potential corresponds to that which would have been generated by signal on the detector; i.e. some level of discharge. This initial potential is then used to calculate electric fields which induce a charge distribution in adjacent pixels and corresponding potentials on neighboring indium bumps. This change in integrated charge on the neighboring pixel with a corresponding change in electrostatic potential on the central indium bump bond yields a capacitance curve of behavior:

\begin{equation}
\label{eq:Cap_IPC}
C_{IPC}=\frac{dQ}{dV} \approx \frac{{\Delta}Q}{{\Delta}V}=\frac{Q_i-Q_{i-1}}{V_i-V_{i-1}}\ .
\end{equation}
In this way our simulations step through voltage on the central pixel (V) to calculate charge on the adjacent pixel (Q) which generates capacitance ($C_{IPC}$). Using equation \ref{eq:coup}, this interpixel capacitance is compared to a nodal capacitance to yield a coupling coefficient. 

The particular diode structure modeled here is a p on n diode\cite{Antoni00} with the substrate removed as is typical of HgCdTe arrays such as those to be used on the James Webb Space Telescope (JWST). The thinned active detector region consists of a 2$\mu$m thick implant of $p=10^{18}cm^{-3}$ HgCdTe embedded with 6$\mu$m of $n=10^{15} cm^{-3}$ for each diode. The implant is in direct contact with a pixel centered cylindrical indium bump of particular initialized diameter. The volume around the indium bump between the HgCdTe and readout multiplexer contains a material defined by an initialized dielectric constant. This pixel structure is replicated to form a three by three grid with 18$\mu$m center to center distance between nearest neighbors.  These values are not all exact; they are best guesses from available information.  Doping profiles, indium bump shapes and sizes, and exact implant sizes among other parameters are proprietary and could not be obtained exactly.  Deviation between true values and estimates is expected to be minimal.
\section{Cosmic Ray Event Exposures}
\label{sec:CRE}
Primary Cosmic rays are high-energy emissions from the solar system and beyond. They are composed primarily of protons with some heavier ions. When these particles collide with atoms in the upper atmosphere they generate a cascade of secondary particles composed largely of pions with a low percentage of kaons. Near the bottom of the atmosphere genuine cosmic rays consist almost exclusively of relativistic muons and neutrinos produced by secondary meson decay\cite{Groom04}. When a muon with high energy strikes a pixel it can generate an immense number of electron-hole pairs. This causes the pixel to suddenly reach an extremely high signal level, in some instances, sufficient to immediately saturate. Depending on the incident angle and energy level, the muon may strike through multiple pixels and create a cluster of bright pixels. At ground level, the muon interactions can be used to measure electrical crosstalk. We have estimated the coupling amount between neighboring pixels by identifying isolated single cosmic ray events and analyzing the signal level in adjacent pixels.
\begin{figure}
\begin{center}
\begin{tabular}{c}
\includegraphics[height=7.6cm]{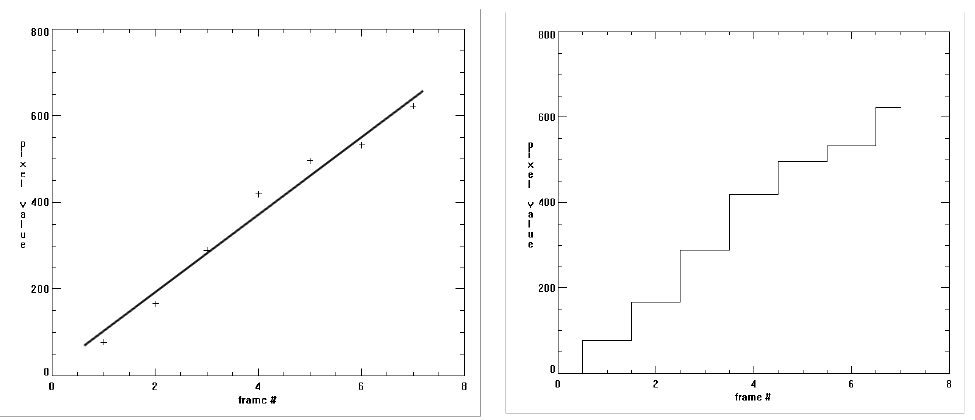}
\end{tabular}
\end{center}
\caption {(left) Up-the-ramp samplings for a typical pixel without incident signal and fit line (right) data rendered in histogram mode.} 
\label{fig:5}
\end{figure} 

In astronomical applications, sampling up-the-ramp is one of the widely-used readout techniques. In this technique, the signal levels are read continuously, non-destructively, and repeatedly at the same temporal frequency for the duration of integration. This readout is then fit to a straight line, the slope of which is proportional to the rate of incident flux. Cosmic ray events can be efficiently identified in the fitting process as an anomalous jump in pixel value. The identified independent cosmic events can be used to measure the IPC magnitude. For a typical pixel, the signal value in a sequence of frames appear to form an approximately straight line. This is shown in figure~\ref{fig:5}; where there are eight non-destructive samplings and the first value has been subtracted from each subsequent sample.
\begin{figure}
\begin{center}
\begin{tabular}{c}
\includegraphics[height=7.6cm]{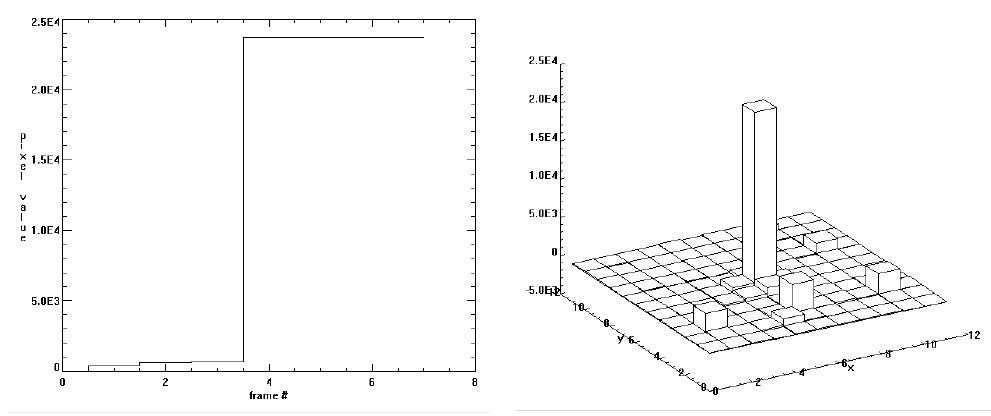}
\end{tabular}
\end{center}
\caption {(left) Up-the-ramp samplings for a typical pixel struck by a CRE rendered in histogram mode. (right) Pixels in the neighborhood of the pixel hit by CRE as a difference frame.} 
\label{fig:6}
\end{figure} 

However, when a cosmic ray strikes a pixel the sampling value will appear quite different. Figure~\ref{fig:6} (left) shows the up-the-ramp sampling of a pixel hit by a typical cosmic event. The pixel signal will jump to a high level suddenly and then either saturate, or resume the prior rate of change for the remainder of the exposure. From the dramatic change of signal level in the pixel, we can determine when the cosmic event hits the detector among the readout sequence. Subtracting the preceding frame from the frame when the cosmic event occurred, a distribution of pixel values around the pixel hit by the cosmic event would be obtained, as seen in figure~\ref{fig:6} (right). The distribution of signals can be used to estimate capacitive coupling.
\subsection{Measuring IPC from CREs}
To utilize single pixel CREs to measure IPC a method must be used which will discern CREs from other high level pixels such as high dark current "hot" pixels.

A plot of all the samplings in a hot pixel during an integration should look akin to that in figure 5, with the exception that the increment amount in the sequence is substantially larger (i.e. the slope is greater). The procedure to measure the IPC from CREs is as follows: 
\begin{enumerate}
\item Identify both potential hot pixels and cosmic events based on a 3${\sigma}$ threshold above the noise floor.
\item Identify all the potential hot pixels according to consistent large slope of pixel levels during an integration and omit them. 
\item Identify non-isolated single cosmic ray events. These are events where their neighbors also include potential cosmic events or hot pixels. Omit non-isolated events.
\item Calculate IPC magnitude using four nearest neighbors surrounding each isolated cosmic ray event through:
\begin{equation}
\label{eq:CRE_Coup}
{\alpha}={\frac{{\sum_{i=0}^{3} {S_{i}}}}{4S_{center}-{\sum_{i=0}^{3} {S_{i}}}}}\ ,
\end{equation}
where $S_{center}$ is the signal on the central pixel and $S_i$ is the signal on the ith nearest neighbor.
\end{enumerate}
This technique to measure $\alpha$ does not provide an unambiguous determination of IPC. Diffusive cross talk is also included in this measurement. Diffusion occurs when a charge carrier is generated in one pixel, but through a series of quasi-random motions and collisions, is collected in an adjacent pixel. Therefore, the coupling parameter $\alpha$ determined from these CRE measurements is an upper bound on the true IPC only coupling value.
\section{Modeling and CRE exposure Results}
\label{sec:Results}
In this section, the results from CRE exposures and simulations will be compared over various parameters. Unless otherwise noted the base parameters are as presented in table~\ref{tab:param}.

\begin{table}[h]
\caption{Parameters used for simulations} 
\label{tab:param}
\begin{center} 
\begin{tabular}{|l|l|} 
\hline
\rule[-1ex]{0pt}{3.5ex} Parameter & Value \\
\hline\hline
\rule[-1ex]{0pt}{3.5ex} Pixel pitch & 18 $\mu$m \\
\hline
\rule[-1ex]{0pt}{3.5ex} Implant width & 13.5 $\mu$m \\
\hline
\rule[-1ex]{0pt}{3.5ex} Implant depth & 2 $\mu$m \\
\hline
\rule[-1ex]{0pt}{3.5ex} Implant carrier concentration & $10^{18} cm^{-3}$ \\
\hline
\rule[-1ex]{0pt}{3.5ex} Bulk thickness & 8 $\mu$m \\
\hline
\rule[-1ex]{0pt}{3.5ex} Bulk carrier concentration & $10^{15} cm ^{-3}$ \\
\hline
\rule[-1ex]{0pt}{3.5ex} Indium bump thickness & 5 $\mu$m \\
\hline
\rule[-1ex]{0pt}{3.5ex} Indium bump diameter & 11 $\mu$m \\
\hline
\rule[-1ex]{0pt}{3.5ex} Epoxy relative dielectric & 4.4 \\
\hline
\rule[-1ex]{0pt}{3.5ex} Temperature & 77K \\
\hline 
\rule[-1ex]{0pt}{3.5ex} Reverse bias & 0.263V \\
\hline
\rule[-1ex]{0pt}{3.5ex} Voltage range & 0 to 0.263V \\
\hline
\end{tabular}
\end{center}
\end{table} 

The coupling coefficient was established by varying the model over particular sets of parameters; these being signal level, temperature, background level, the presence or absence of epoxy underfill and the indium bump diameter. This allows for comparison with CRE results as well as comparison to conclusions from the earlier work's measurements of Brown and Schubnell\cite{Brown06} and Finger, et al \cite{Finger06}. Additionally, bump diameter scaling allows for comparison to a parallel cylinder model and aids in determining the primary cause of the coupling.

\subsection{Signal intensity}
\begin{figure}
\begin{center}
\begin{tabular}{c}
\includegraphics[height=9.5cm]{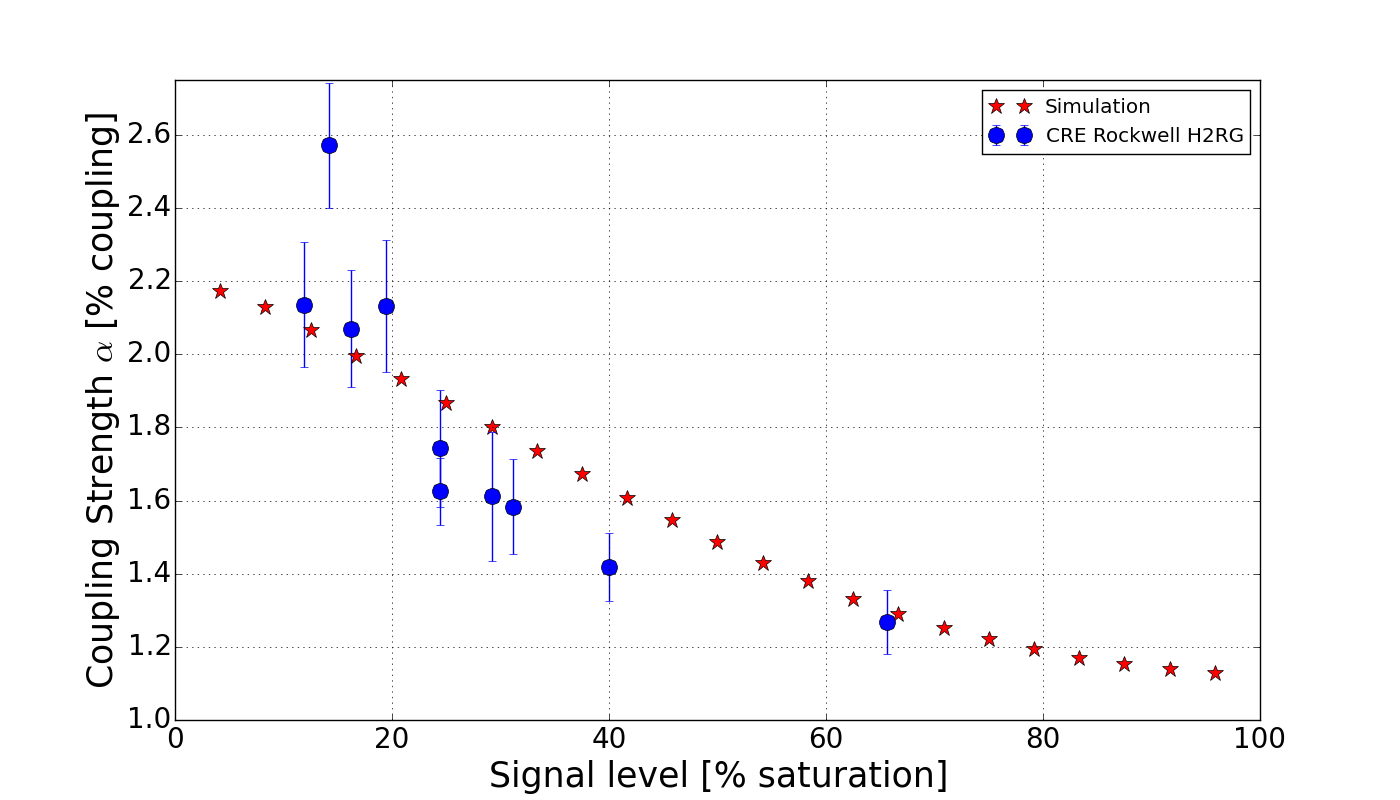}
\end{tabular}
\end{center}
\caption {Coupling as a function of signal strength. Signal is presented as a percentage of depletion of the biasing potential.} 
\label{fig:7}
\end{figure}

The results of modeling are shown in figure \ref{fig:7}; where the coupling coefficient $\alpha$ is plotted as a function of signal level. As a function of relative signal strength, all other factors being equal, the coupling strength is observed to decrease with increasing signal level. This indicates that across an image, a bright pixel will couple less to its neighbors than a dim pixel as a percentage of signal generated. Examining a simulated pixel over the range from no signal to saturation, the average slope of the coupling coefficient was given by
\begin{equation}
\label{eq:sim_slope}
\frac{d\alpha}{dS} \approx -0.0104 \frac { \%\ coupling} {\%\ saturation}.\
\end{equation}
While for CRE data average slope is given by
\begin{equation}
\label{eq:CRE_slope}
\frac{d\alpha}{dS} \approx -0.0507 \frac { \%\ coupling} {\%\ saturation}.\
\end {equation}
A linear characterization is not the most appropriate $(r^2<.81)$. This is seen in figure~\ref{fig:7}, where it is clear that the same change in signal level will result in a larger change in coupling coefficient at smaller signal levels compared to larger signal levels. Both the coupling and the rate of change of the coupling are decreasing with increasing signal strength.

In comparison to CRE data, the simulated data is an underestimate of both peak coupling and minimum coupling with a central region where coupling is overestimated. CRE peak coupling is approximately 2.6\%\ at low signal levels with minimum coupling of 1.275\%\ near saturation. The simulations peak coupling is approximately 2.19\%\ at low signal levels with a minimum coupling of 1.175\%\ near saturation. 

\begin{figure}
\begin{center}
\begin{tabular}{c}
\includegraphics[height=3.3cm]{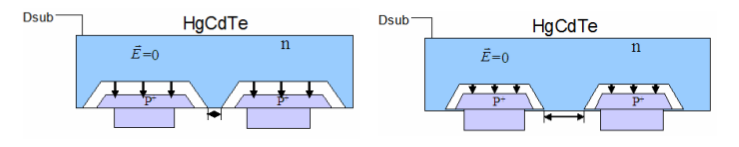}
\end{tabular}
\end{center}
\caption {Depletion region for HgCdTe arrays (left) before charge collection, (right) after charge collection)} 
\label{fig:8}
\end{figure}

This decrease in coupling with stronger signals can be explained by a shrinking of the depletion region which occurs as a larger signal is accumulated on a pixel. This shrinking of the depletion region simultaneously decreases exposed area of the depletion region, increases the distance between adjacent regions, and reduces the peak electric field magnitude within the region as illustrated by figure~\ref{fig:8}. Each of these trends serve to decrease interpixel capacitance and therefore decrease the coupling coefficient.
\subsection{Epoxy}

An epoxy under-fill is regularly added to mechanically and thermally stabilize the hybridization. The epoxy fills the gap between the indium bumps which changes the dielectric constant of this region. This filling amplifies the IPC that occurs between adjacent bumps linearly by 
\begin{equation}
\label{eq:dielectric}
C_{IPC dielectric} = \varepsilon _r C_{IPC vacuum}.\
\end{equation}
Inserting this result into equation~\ref{eq:coup} returns the expected coupling. This increase in interpixel capacitance would yield a non-linear increase in the coupling coefficient given by
\begin{equation}
\label{eq:dielec_coup}
\alpha _{vacuum} = \frac {C_{IPC\ vacuum}}{4C_{IPC\ vacuum}+ C_{node}} \rightarrow \alpha _{dielectric}= \frac{\varepsilon _r C_{IPC\ vacuum}}{4 \varepsilon _r C_{IPC\ vacuum}+ C_{node}}=\frac {C_{IPC\ vacuum}}{4C_{IPC\ vacuum}+ \frac{C_{node}}{\varepsilon _r}}.\
\end{equation}
If relation~\ref{eq:dielectric} holds for $C_{IPC}$ then the coupling coefficient is effectively changed as if the nodal capacitance were decreased by a factor of the dielectric constant.

Utilizing an epoxy with a relative dielectric constant $\varepsilon _r =4.4$ it is found from the models that coupling increases, though not by the amount expected. This indicates that the coupling is not exclusively occurring between adjacent indium bumps as had earlier been postulated. It was suggested by Brown and Schubnell, among others, that this discrepancy may be caused by coupling that occurs within the ROIC as the depleted detector is field free\cite{Brown06}.
\begin{figure}
\begin{center}
\begin{tabular}{c}
\includegraphics[height=10.1cm]{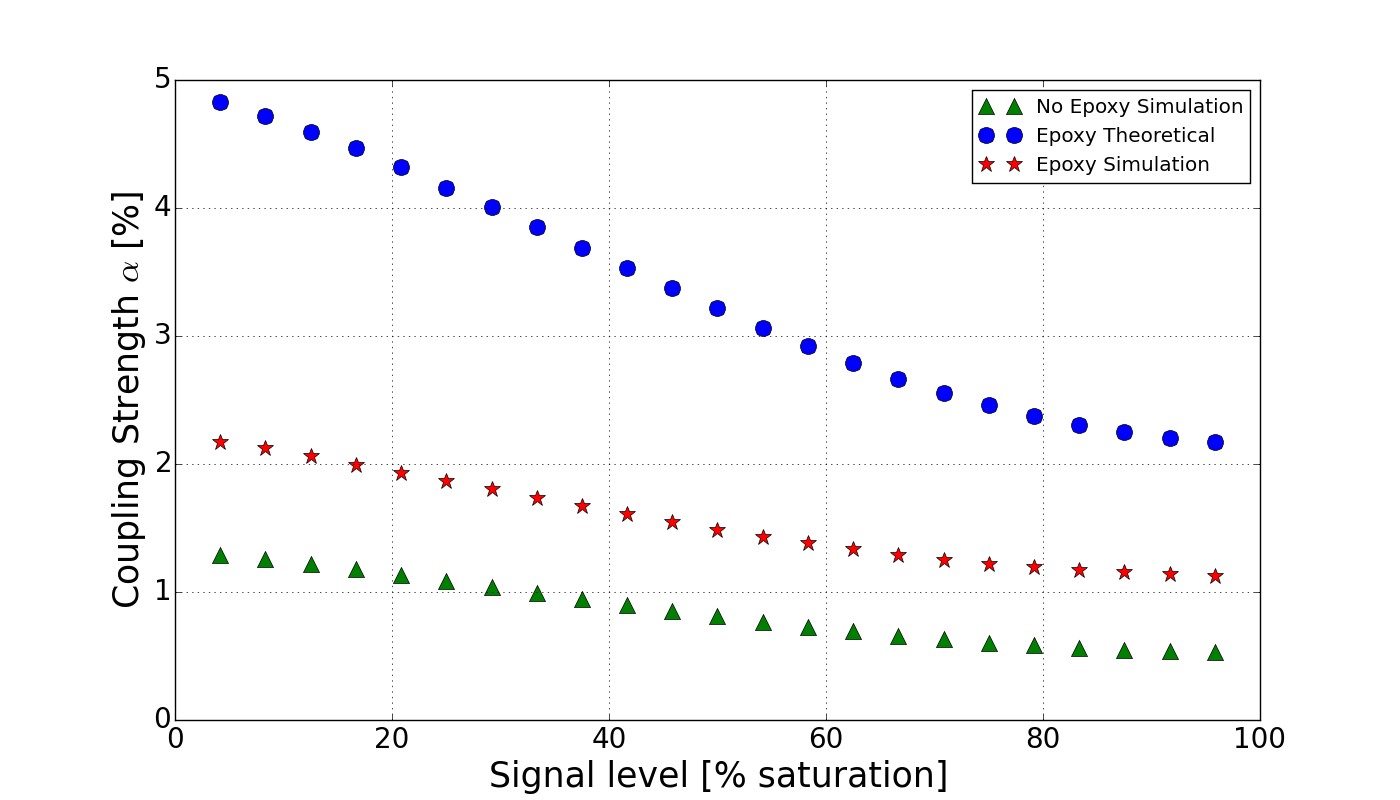}
\end{tabular}
\end{center}
\caption {Simulation results with and without epoxy present as well as a theoretical result from equation~\ref{eq:dielec_coup} utilizing the simulated $C_{IPC\ vacuum}$ values with 4.4 as the dielectric constant. The disagreement between simulated and calculated values indicates that the coupling between indium bump is not the only culprit for IPC.} 
\label{fig:9}
\end{figure} 

Simulations were performed, with results presented in figure~\ref{fig:9}, where the relative dielectric constant of the under-fill was switched to be either 1 or 4.4. The resulting coupling increased by an average factor of 1.87 when the epoxy was present. This is substantially smaller than the 3.95 times increase expected if equation~\ref{eq:dielectric} held true for $C_{IPC}$. This result indicates that the difference between theoretical parallel cylinder coupling and measured coupling cannot be explained by an additive term from within the ROIC, as these models did not include the ROIC but still disagree with parallel cylinder coupling.

A likely explanation for this is that the angle of the electric field coming out of the depletion region is not always perpendicular to the boundary between the depletion region and the dielectric.  By solving Poisson's equation in the presence of bound charge the following relationships at the boundary fall out\cite{Ohanian07}:
\begin{equation}
\label{eq:BVPpar}
E_{1\parallel} =E_{2\parallel}
\end{equation}
\begin{equation}
\label{eq:BVPperp}
\varepsilon_1 E_{1\perp}=\varepsilon_2 E_{2\perp}
\end{equation}
This indicates that only the portion of the electric field perpendicular to the interface is amplified by the dielectric constant.  When fields are at an oblique angle, the effect of changing materials is minimized.  Fields exiting a conductor are necessarily perpendicular to the interface between materials;  this is a constraint which stems from requiring zero magnitude internal electric fields.  Boundaries between insulating materials, such as those that occur at the boundary between the depletion region and the epoxy, are not subject to this constraint.
\subsection{Temperature}
\begin{figure}
\begin{center}
\begin{tabular}{c}
\includegraphics[height=10.1cm]{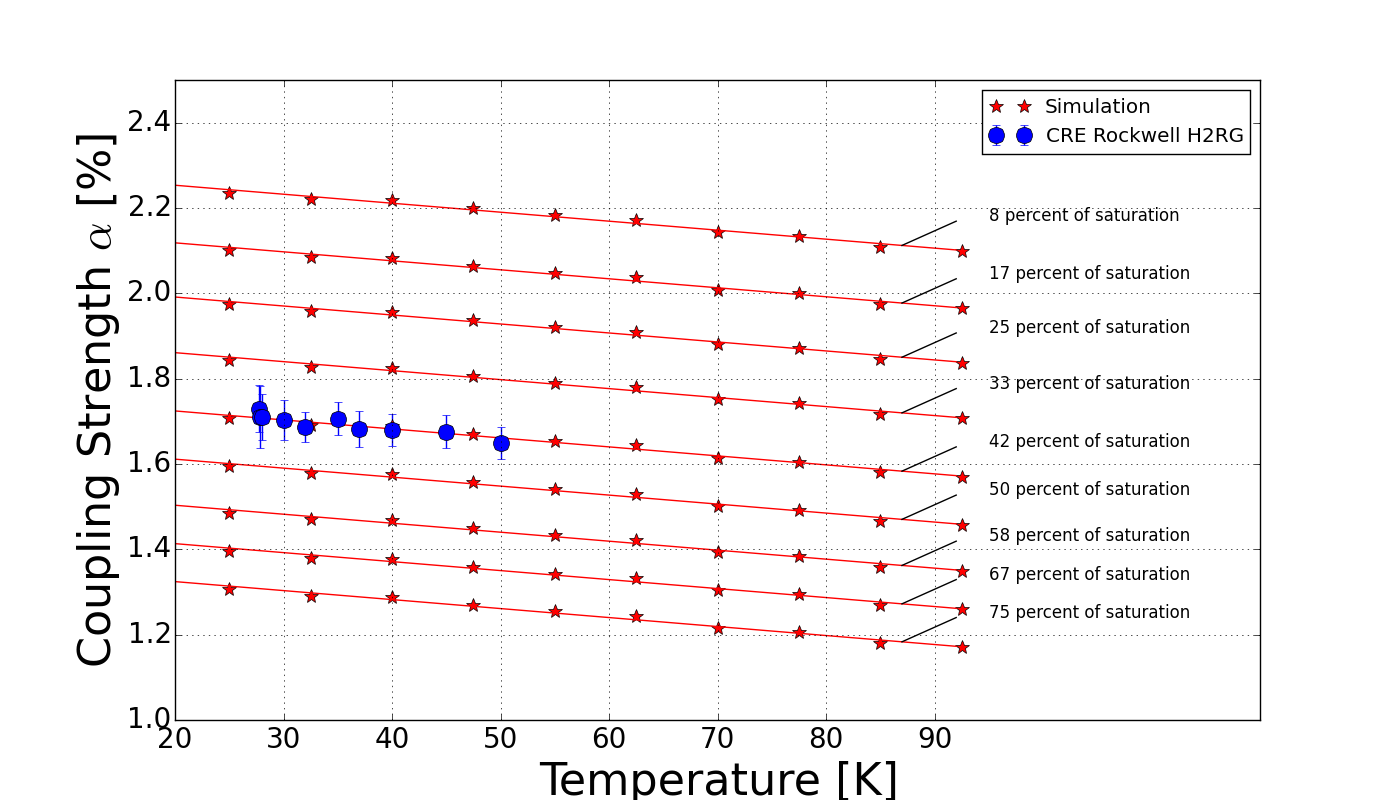}
\end{tabular}
\end{center}
\caption {Coupling as a function of temperature with overlaid linear regressions. The higher coupling regressions are on data with a low signal level while the lower coupling regressions are on data with a higher signal level.} 
\label{fig:10}
\end{figure}

The governing equations~\ref{eq:poisson} -~\ref{eq:dde_4} contain temperature dependence through their diffusivity $(D_{n,p})$ and mobility $(\mu _{n,p})$ terms which impact the statistical recombination rates, mean free paths, and ultimately resistivity and permittivity of the materials. Thus electric field strength has a dependence on temperature. Resistivity of the bulk drops at lower temperatures resulting in a greater mobilization of charge for an equivalent change in electric field. This is in opposition to increasing permittivity of the bulk material with increasing temperature. The observed result from both CREs and simulations of all temperature dependence is a decreasing coupling with increasing temperature as presented in figure~\ref{fig:10}.
Taking the average slope of these curves yields
\begin{equation}
\label{coup_temp}
\frac{d\alpha }{dT} \approx -0.0021 \frac {\%}{K}\ .
\end{equation}
This temperature effect is negligibly small compared to the change in IPC attributed to changing signal level. For example, the NIRcam device for the JWST will have temperature change during operation $\leq $ 0.1K\cite{Ozborn07} which would result in a change in coupling of approximately 0.00021$\% .$ In comparison to result from section 4.1, saturation over a full range would be expected to change 1.015$\% $, a change four orders of magnitude larger. This indicates that relative to the other examined parameters, variation of coupling coefficient due to temperature change can be safely ignored under expected operating conditions for most sensor systems.
\subsection{Background intensity}
\begin{figure}
\begin{center}
\begin{tabular}{c}
\includegraphics[height=10.1cm]{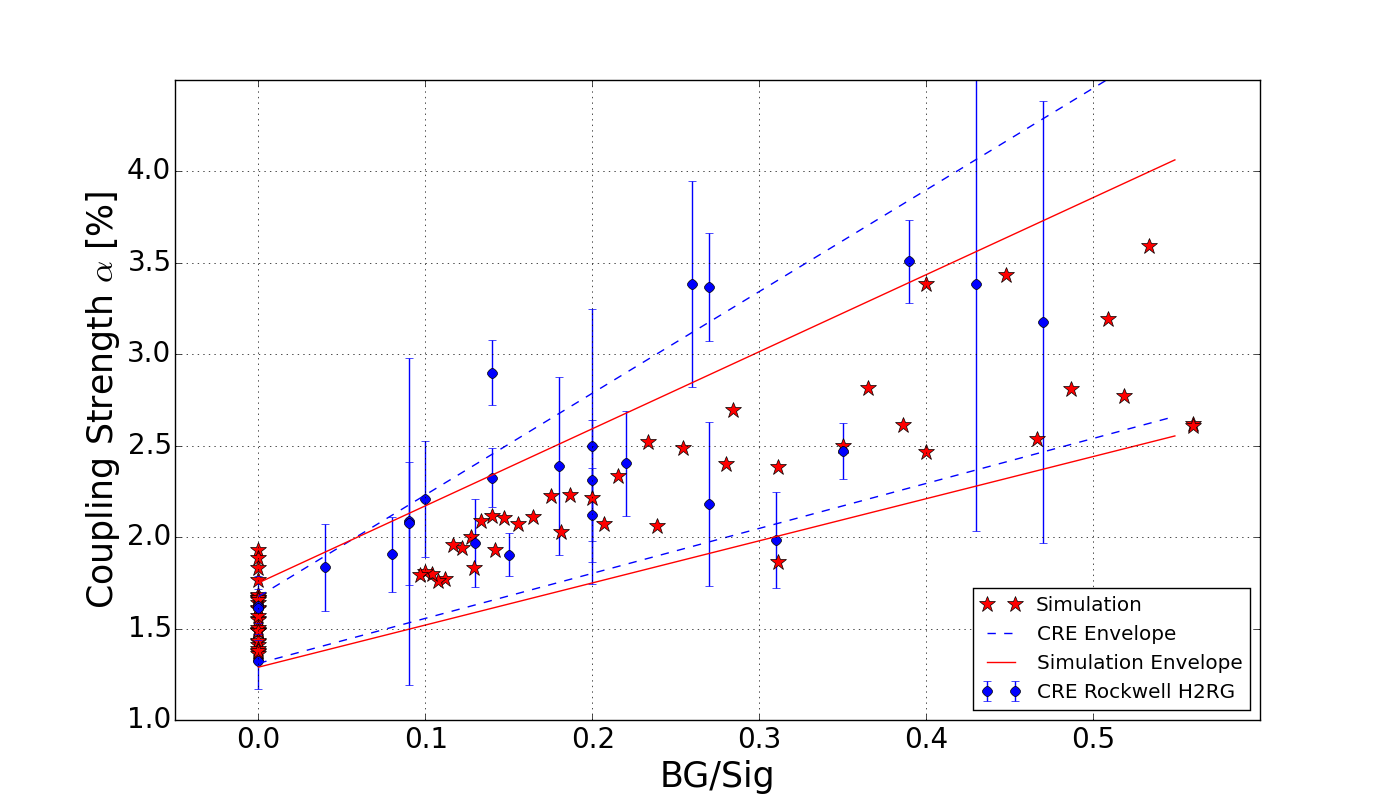}
\end{tabular}
\end{center}
\caption {Comparison between CRE measured coupling and simulated coupling as a function of the ratio between background and signal intensity.} 
\label{fig:11}
\end{figure}

As a function of the ratio between signal and background level, all other factors being equal, the coupling strength is observed to increase. This taken in tandem with the dependence on signal strength results in both a higher coupling with higher relative background to signal as well as a higher spread in coupling as presented in figure~\ref{fig:11}. This is not a functional relationship and gives a higher spread due to the plurality of signal levels that can yield the same ratio.
By approximating linear envelopes over and under this data the following relationships are obtained.
\begin{equation}
\label{eq:bg_cre_up}
\alpha _{upper\ measured} [\%] = 5.5597 \left( \frac{BG}{Sig} \right) + 1.675,\
\end{equation}
\begin{equation}
\label{eq:bg_cre_low}
\alpha _{lower\ measured} [\%] = 2.4592 \left( \frac{BG}{Sig} \right) + 1.310,\
\end{equation}
\begin{equation}
\label{eq:bg_sim_up}
\alpha _{upper\ simulated} [\%] = 4.2134 \left( \frac{BG}{Sig} \right) + 1.750,\
\end{equation}
\begin{equation}
\label{eq:bg_sim_low}
\alpha _{lower\ simulated} [\%] = 2.3031 \left( \frac{BG}{Sig} \right) + 1.289.\
\end{equation}
This behavior indicates that both the absolute signal level and the relative signal and background levels can impact coupling. With a fixed signal level, an increasing background level monotonically increases coupling. Similarly, with a fixed background level, an increasing signal level monotonically decreases coupling. Equations~\ref{eq:bg_cre_up} -~\ref{eq:bg_sim_low} then provide an approximate “best-case” and “worst-case” for an expected relative background and signal strength for this detector.

In comparing the simulated and measured data presented in figure~\ref{fig:11}, the enveloping functions for CREs has a higher slope for both the upper and lower bounds.

Due to practical memory and timely convergence constraints in the simulations, a uniform $\Delta V$ was applied to the central and neighboring pixels.  This results in the simulated data points with low $\left( \frac{BG}{Sig}\right)$ ratio also have high signal magnitude, while the points with higher $\left( \frac{BG}{Sig}\right)$ ratio have a wider range of signal values. This results in points at low $\left(\frac{BG}{Sig}\right)$ values on figure~\ref{fig:11} having lower coupling because the data points are weighted with high signal. Simultaneously during CRE detection, to meet the $3\sigma$ criterion for CRE event identification at higher background levels, a larger signal value in a pixel would be required to be noticed above background.

Though trends are visible, the full nature of the impact of background level on coupling remains unclear.  The extent of the conclusions that can be drawn here are that a higher background level seems to cause a higher coupling. Additionally, the highest measured couplings are observed to occur with high backgrounds and low signals. These couplings top at 3.5\%\ for CRE data and 3.6\%\ in simulations.
\subsection{Indium bump diameter}

\begin{figure}
\begin{center}
\begin{tabular}{c}
\includegraphics[height=3.5cm]{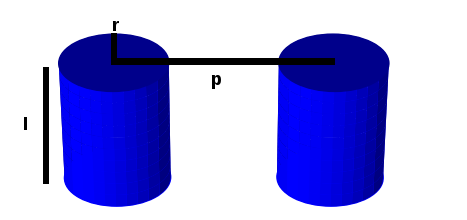}
\end{tabular}
\end{center}
\caption {Configuration where capacitance is governed by equation~\ref{eq:par_cyl_cap}. Two parallel conductive cylinders of equal radius (r), length (l) and center-to-center distance (p). } 
\label{fig:added}
\end{figure}

As the indium bumps increase in size, with a constant pixel pitch, the distance between them decreases. By modeling the indium bumps as conductive cylinders the expected capacitance between them would be:
\begin{equation}
\label{eq:par_cyl_cap}
C=\varepsilon l \int^{r}_{-r}\frac{1}{p\sqrt{1-\frac{x^2}{r^2}}} dx = \frac{\pi \varepsilon l}{{arcosh} \left( \frac{p}{2r} \right)}.\
\end{equation}
Where $\varepsilon$ is the electric permittivity of the dielectric material between, l is the length of the cylinders, r is their radius, p is the center to center distance between the cylinders, or in our case, the pixel pitch, and arcosh is the inverse hyperbolic cosine function. The geometric parameters are as presented in figure ~\ref{fig:added}.

\begin{figure}
\begin{center}
\begin{tabular}{c}
\includegraphics[height=6.8cm]{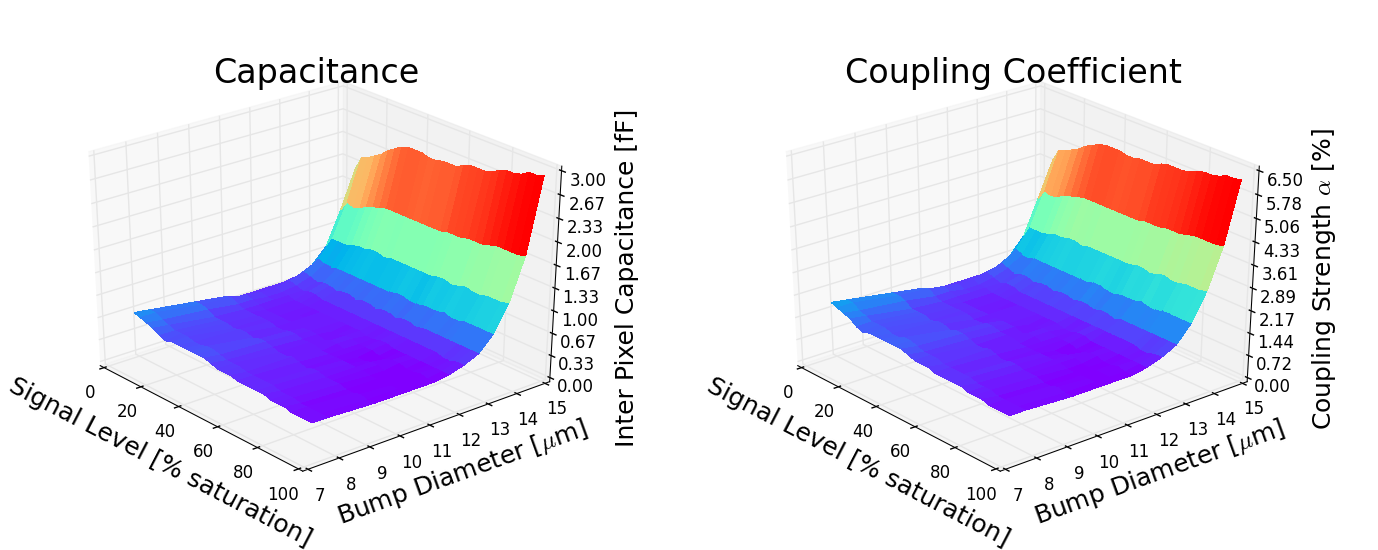}
\end{tabular}
\end{center}
\caption {Simulated IPC between adjacent pixels as a function of both signal intensity and indium bump diameter as (left) capacitance and (right) coupling coefficient with assumed nodal capacitance of 33.5 fF} 
\label{fig:12}
\end{figure}

In simulations bump diameters were varied from 7.5$\mu$m to 15$\mu$m while the pixel pitch was held constant at 18$\mu$m. Results are presented in figure~\ref{fig:12}.
When the distance between the bumps becomes small (i.e. when the bumps themselves become large) the coupling magnitude skyrockets and the behavior of the IPC is approximately that of the parallel cylinder capacitor. The expected behavior of the coupling if it were exclusively due to coupling between indium bumps would be that from equation \ref{eq:par_cyl_cap}.

To ease examination of this behavior, coefficients of potential can be examined instead, which are defined as the inverse of the capacitance.
\begin{equation}
\label{eq:Cof_Pot}
P \equiv \frac{1}{C} = {k} \ {arcosh} \left( \frac{p}{2r} \right).\
\end{equation} 
With a pixel pitch of 18$\mu$m as the diameter scales from 7.5 to 15 $\mu$m this function is nearly linear as a function of bump radius.
\begin{figure}
\begin{center}
\begin{tabular}{c}
\includegraphics[height=9.0cm]{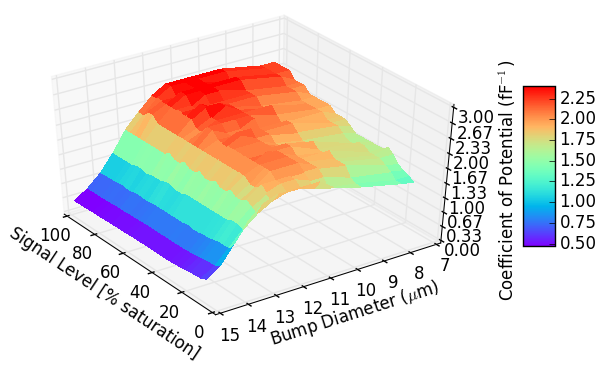}
\end{tabular}
\end{center}
\caption {Coefficients of potential as bump diameter is scaled from 7.5 to 15 $\mu$m.} 
\label{fig:13}
\end{figure} 

\begin{figure}
\begin{center}
\begin{tabular}{c}
\includegraphics[height=6.4cm]{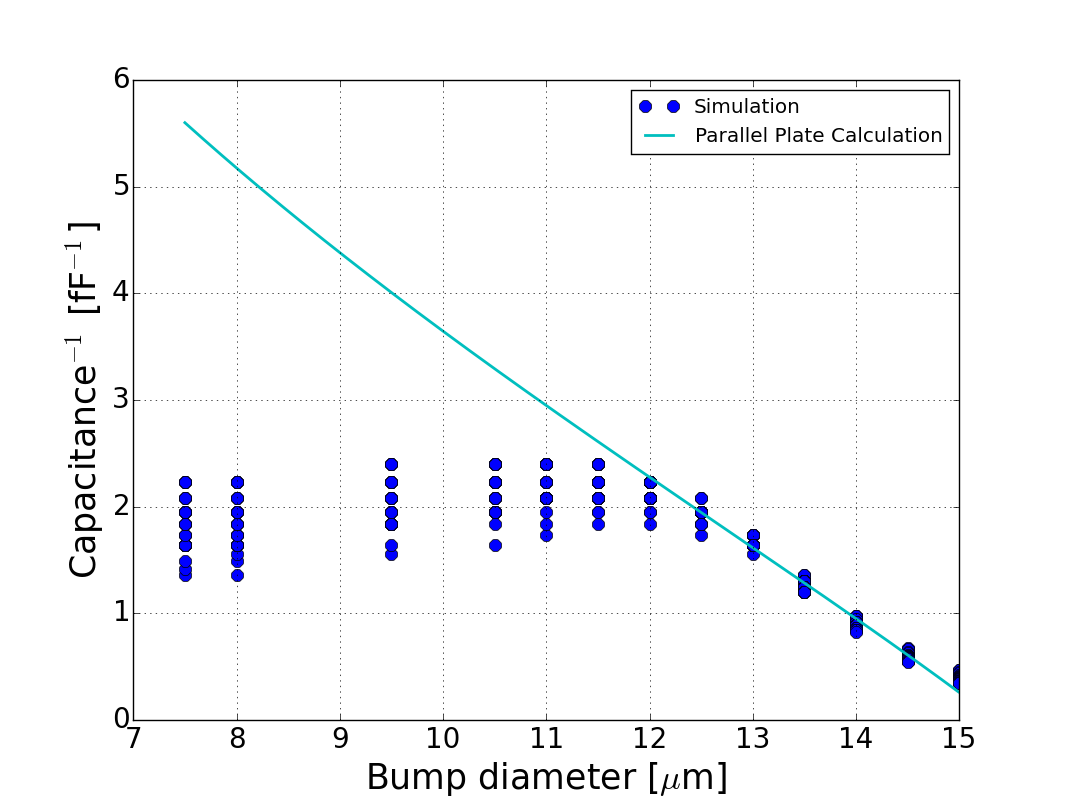}
\end{tabular}
\end{center}
\caption {Comparison between the simulated data and predicted results if coupling were occurring exclusively as a parallel cylinder capacitor between indium bumps for an 18$\mu$m pixel pitch as bump diameter scales from 7.5 to 15$\mu$m.} 
\label{fig:14}
\end{figure}

This situation corresponds to the circumstance where it is appropriate to describe IPC as dominated by the capacitive coupling between adjacent indium bump bonds. Figures~\ref{fig:13} and~\ref{fig:14} illustrate that the coupling can be taken as exclusively between the bumps only when the bump diameter is greater than 13$\mu$m. This corresponds to a separation between the bumps of less than 5$\mu$m for a pixel pitch of 18$\mu$m. When the separation is greater than 5$\mu$m the coefficient of potential no longer behaves linearly which indicates that the coupling cannot adequately be described as the coupling between indium bumps.
\subsection{Coupling fields}
\begin{figure}
\begin{center}
\begin{tabular}{c}
\includegraphics[height=8.89cm]{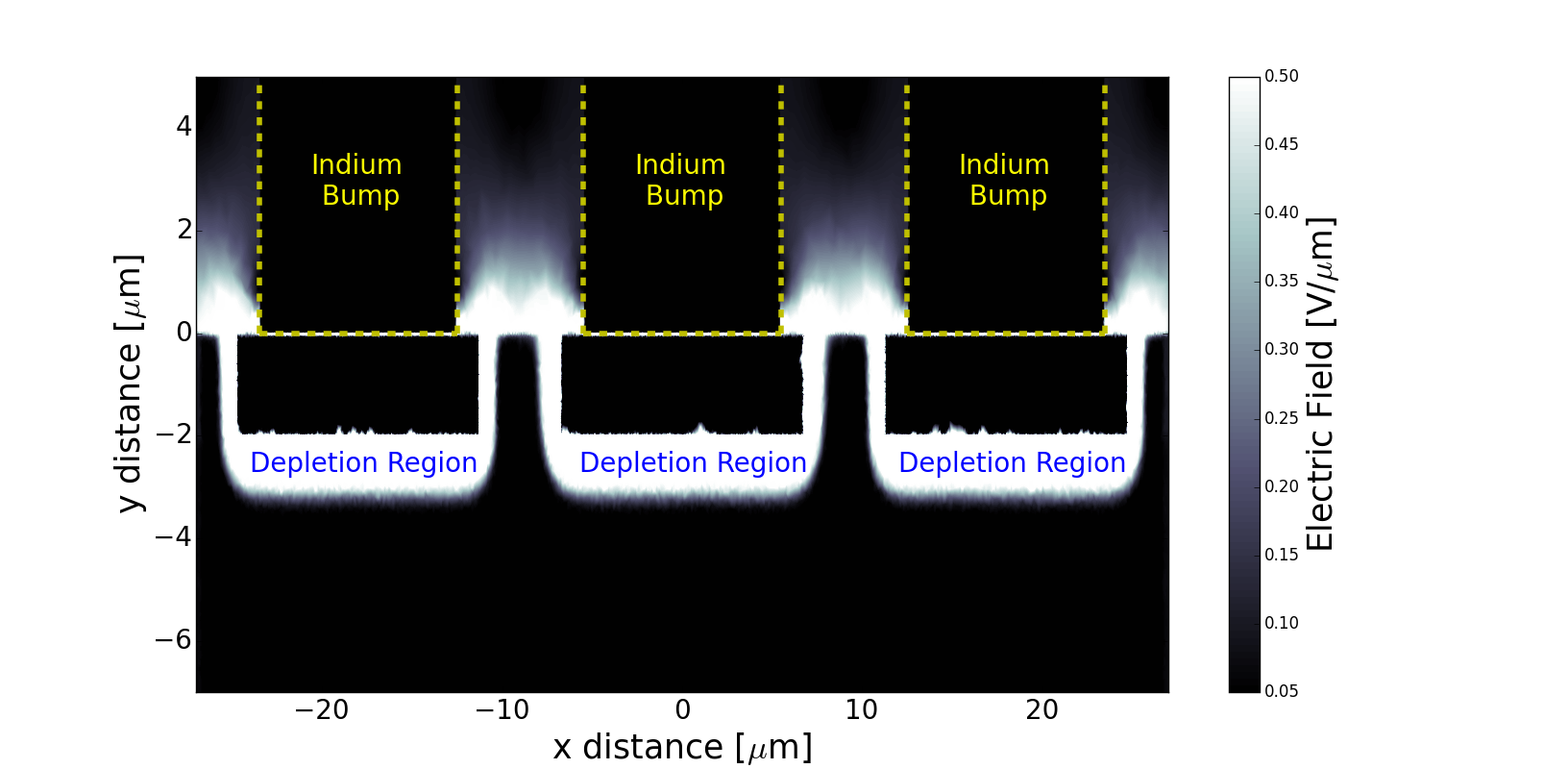}
\end{tabular}
\end{center}
\caption {Electric field cross-section through the array in the case where no discharge has occurred (zero signal). Note the zero magnitude electric fields in the bulk, implant, and indium. Fringing fields are visible where the depletion region contacts the epoxy. The high fields of the depletion region are display clipped to allow for observation of fringing fields.} 
\label{fig:15}
\end{figure}

Figure~\ref{fig:15} illustrates the electric field distribution in the detector as generated by the simulation software. By examining two dimensional electric field magnitude cross sections, certain facets of the coupling can be elucidated. Within the HgCdTe bulk, the intrinsic regions posess approximately zero magnitude electric field resulting in the depletin regions being the only volumes where significant electric fields can exist.  As the reverse bias is discharged the size and location of the depletion regions change as well as the magnitude of electric field within each depletion region. Where the depletion regions meet the dielectric under-fill, strong fields pervade out; these are referred to as the fringing fields. Within the indium/epoxy layer the only region where non-zero electric fields can exist is within the epoxy, as the indium is a conductor and therefore supports no internal electric fields. The only net charge that can exist within the indium is as a surface charge distribution which is responsible for external fields and internal potential.

Figure~\ref{fig:15} clearly shows the presence of these fringing fields originating at the boundary between the depletion region and the dielectric epoxy. The nearer to the depletion region the stronger the fields. These fields permeate out until they contact the indium bumps where they adjust the surface charge distribution and influence the electrostatic potential of the contact.
\begin{figure}
\begin{center}
\begin{tabular}{c}
\includegraphics[height=8.50cm]{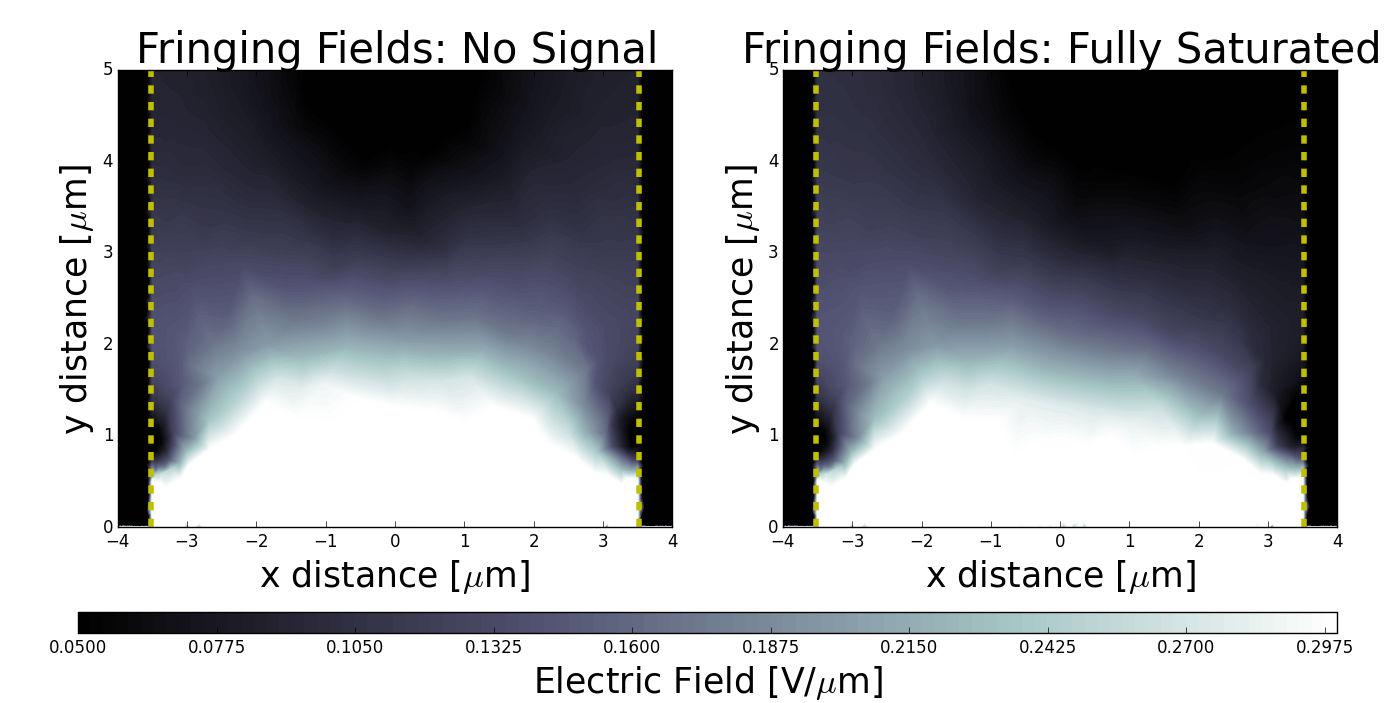}
\end{tabular}
\end{center}
\caption {The region of the fringing fields in two cases: (left) when no signal is incident on either pixel, (right) when the left pixel has had no signal but the right pixel is in a state of complete discharge/saturation. High fields are display clipped to allow for observation of fringing fields throughout the epoxy.} 
\label{fig:16}
\end{figure} 

\begin{figure}
\begin{center}
\begin{tabular}{c}
\includegraphics[height=8.89cm]{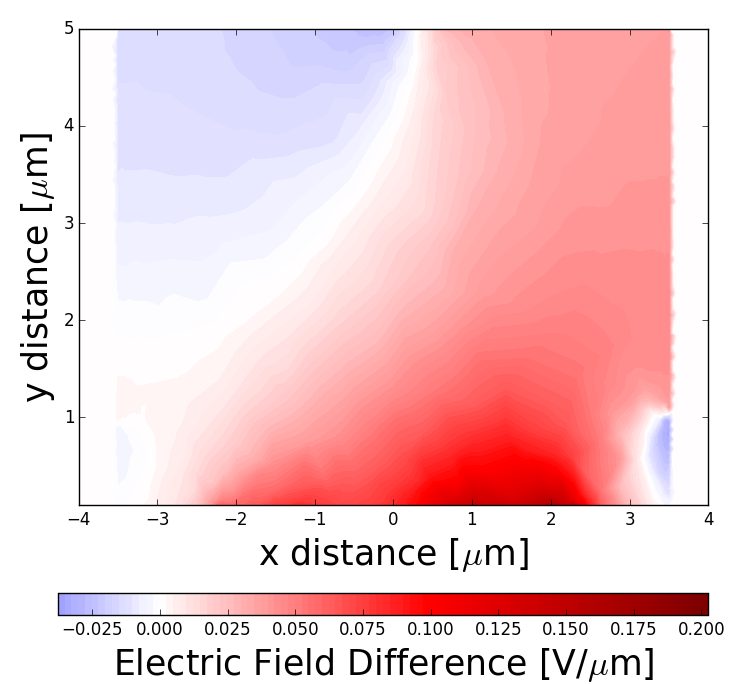}
\end{tabular}
\end{center}
\caption {A difference plot between the two cases presented in figure 17. This case is (no signal - saturated). The electric fields are greater in magnitude during the undepleted case. Signal collection discharges the diode resulting in a smaller depletion region with weaker fields. The negative field in near the right indium bump is due to the change field difference of a single diode as the bump is discharged.} 
\label{fig:17}
\end{figure} 

Examining the fringing fields at the extremes of signal detection in figure~\ref{fig:16} shows an asymmetry. The field fringing from the pixel which has discharged is weaker in magnitude than its still fully charged counterpart. This results in the potential from the discharged pixel having a weaker connection to its still charged neighbor than in the case where both pixels remain fully charged. The magnitude of this fringing field dominates the field due to indium bumps being at different electrostatic potentials. This phenomenon can be better illustrated as a difference plot as seen in figure~\ref{fig:17} ,where it is more clear that through-out the majority of the under-fill region the electric field strength is higher in the fully charged state compared to the discharged state. The two excepting regions are the portion furthest from the depleted diode, where the fringing field is weakest; and the region directly above the depleted diode.  In the top right region, the electric field due to the difference in electrostatic potential between the indium bumps is dominating.  The direction of this field is in opposition to the direction of fields from the depletion region causing a region where the field decreases. The region in the bottom right is where the contact between the bump and diode largely shield from the fringing fields.  This again allows the parallel fields to contribute significantly in opposition to the fringing fields.

The reason for the change in fringing field magnitude is that as a pixel discharges the length of its depletion region decreases. This decrease is due to a decrease in potential difference across the diode. Simultaneously, this decrease in potential difference corresponds to a decrease in field magnitude. Together, these serve to weaken the strength of the field both within the depletion region and which fringe out from the depletion region, resulting in a lower coupling.

Observation of these fields gives insight into a suggested mechanism of eliminating this portion of coupling by absorbing these fields. Introduction of an additional electronic contact capping the depletion regions on the other side of the passivation layer would attenuate this cross-talk. Grounding this cap would serve to greatly diminish and potentially eliminate this aspect of coupling.
\section{Conclusion}
In order to give additional context to these simulations the IPC results can be compared to earlier reported values from literature as presented in table~\ref{tab:lit}. Any capacitive coupling introduced in the ROIC would be introduced in parallel to coupling in the indium interconnect layer and sensor layer. This would result in an additive capacitance. Because the ROIC was not simulated it is expected that compared to measured sensor IPC with the same physical parameters the modeled capacitance will be a lower bound. Furthermore, historical IPC results for similar 18$\mu$m sensors are within the range presented by these simulations.

\begin{table}[h]
\caption{List of IPC value both measured and presented from literature. In cases where IPC was not given directly but instead a nodal capacitance and coupling were reported, IPC was calculated from equation~\ref{eq:coup}. Row vs column and fast vs slow refer to multiplexer directions.} 
\label{tab:lit}
\begin{center} 
\begin{tabular}{|l|l|l|} 
\hline
\rule[-1ex]{0pt}{3.5ex} Source & IPC & Reported IPC coupling \\
\hline\hline
\rule[-1ex]{0pt}{3.5ex}HgCdTe SB-301\cite{Brown06} & 0.41449 fF or 0.16668 fF & 0.54\%\ or 0.22\%\ \\ no underfill &(rows vs. columns)& \\ (20 $\mu$m pitch) & & \\
\hline
\rule[-1ex]{0pt}{3.5ex}HgCdTe SB-301\cite{Brown06} & 1.0224 fF or 0.38026 fF & 1.25\%\ or 0.48\%\ \\ underfill &(rows vs. columns)& \\ (20 $\mu$m pitch) & & \\
\hline
\rule[-1ex]{0pt}{3.5ex}Rockwell HgCdTe\cite{Brown06} & 0.765155 fF & 2.17\%\ \\ 1.7$\mu$m cutoff & & \\ H2RG with epoxy& & \\ (18 $\mu$m pitch) & & \\
\hline
\rule[-1ex]{0pt}{3.5ex}HgCdTe\cite{Finger06} & 0.63038fF or 0.51565 fF & 1.75\%\ or 1.45\%\ \\ 2.5$\mu$m cutoff &(fast vs. slow)& \\ H2RG with epoxy & & \\ (18 $\mu$m pitch) & & \\
\hline
\rule[-1ex]{0pt}{3.5ex}Measured HgCdTe & 0.4408 fF to 1.363 fF & 1.25-3.5\%\ \\ H2RG using CREs & & \\ 5 $\mu$m cutoff & & \\ in this paper & & \\
\hline
\rule[-1ex]{0pt}{3.5ex}Result from simulations & 0.4038 fF to 1.408 fF & 1.15-3.6\%\ \\ in this paper & & \\
\hline
\end{tabular}
\end{center}
\end{table}

The generated simulations corroborate trends observed in the CRE data which indicate that the coupling strength is not constant. The IPC is dependent on signal strength, background strength, and to a lesser degree, temperature. As a result, simple filtering techniques are insufficient to restore photometric accuracy to images with varying signal levels across the scene. The modeled behavior of IPC is in agreement with the measured behavior of IPC as a function of these parameters.

Further work must be done on the characterization of IPC, particularly as a function of background signal level. This requires both additional simulation and more controlled measurements. In particular, measurements of IPC over a range of signal strength when fixed above a various background levels should be made.  In order to achieve this, measurements using a single pixel reset method in tandem with a background flux will be utilized.  Simultaneously, the impact of non-constant coupling on scientific imaging is being explored.

The aforementioned results indicate that the situation where the highest coupling occurs is when observing a field in which the signal level is small and background level is large. This implies that IPC will have the largest effect in the situation where the signal is already the most difficult to detect.

\acknowledgments 
The research described in this paper was done by the authors with financial support by NASA through contract NAS5-02105. Simulations were run on the BlueHive2 computing cluster supported by the University of Rochester Center for Integrated Computing with a faculty affiliation through Craig McMurtry. We thank both Craig McMurtry and the University of Rochester Research Computing staff for their assistance.


\bibliography{./report.bib} 
\bibliographystyle{./spiejour.bst} 


\vspace{2ex}\noindent{\bf Kevan Donlon} is a PhD student at Rochester Institute of Technology's College of Science in the Chester F. Carlson Center for Imaging Science.  He received his BS degree in 2012 from Rensselaer Polytechnic Institute's College of Science in physics.
\vspace{2ex}
\\
\noindent{\bf Zoran Ninkov} is a professor at the Rochester Institute of Technology.  He recieved his BSc 1st class Honors in physics at the University of Western Australia in 1978, his MSc degree in physical chemistry from Monash University in 1981, and his PhD in astronomy from the University of British Columbia.  His research interests are in the area of instrumentation development, detector testing, and MEMS devices.
\vspace{2ex}
\\
\noindent{\bf Stefi Baum} is the Dean of the Faculty of Science and Professor of Physics and Astronomy at the University of Manitoba.  She earned a BA in physics with honors from Harvard University and a PhD in astronomy from the University of Maryland.  Her personal research focuses in two areas, the study of activity in galaxies and its relation to galaxy evolution, and the development of image processing, statistical algorithms, and calibration techniques for brain imaging for the diagnosis of mental health and learning diabilities.
\vspace{2ex}
\\
\noindent{\bf Linpeng Cheng} obtained earned is BS in Astronomy from Beijing Normal University in 1999, an MS in Astrophysics from the Chinese Academy of Sciences in 2002, and an MS in Imaging Science from Rochester Institute of Technology in 2009.  He is currently employed by OmniVision Technologies.

\end{spacing}

\end{document}